# High-Energy Calibration of a BGO detector of the GLAST Burst Monitor

Andreas von Kienlin[*], Gerald J. Fishman[¶], Michael S. Briggs[§], Gary L. Godfrey[†] and Helmut Steinle[*]

[*]*Max-Planck-Institut für extraterrestrische Physik, Giessenbachstraße, 85748 Garching, Germany*
[¶]*Marshall Space Flight Center, VP62, Huntsville, AL 35812, USA*
[§]*University of Alabama, NSSTC, 320 Sparkman Drive, Huntsville, AL 35805, USA*
[†]*Stanford Linear Accelerator Center, 2575 Sand Hill Road, Menlo Park, CA 94025, USA*

**Abstract.** The understanding of the instrumental response of the GLAST Burst Monitor BGO detectors at energies above the energy range which is accessible by common laboratory radiation sources (< 4.43 MeV), is important, especially for the later cross-calibration with the LAT response in the overlap region between ~ 20 MeV to 30 MeV. In November 2006 the high-energy calibration of the GBM-BGO spare detector was performed at the small Van-de-Graaff accelerator at SLAC. High-energy gamma-rays from excited $^8$Be* (14.6 MeV and 17.5 MeV) and $^{16}$O* (6.1 MeV) were generated through (p, γ)-reactions by irradiating a LiF-target. For the calibration at lower energies radioactive sources were used. The results, including spectra, the energy/channel-relation and the dependence of energy resolution are presented.

**Keywords:** Instruments: **G**LAST, GBM; calibration
**PACS:** 95.55.Ka, 29.40.Mc, 98.70.Rz

## CALIBRATION MEASUREMENTS & RESULTS

The NaI(Tl)- and BGO-detectors [1] of the GLAST Burst Monitor (GBM) [2] were extensively calibrated with radioactive sources in laboratory measurements [3]. In the energy range of the BGO-detectors from 150 keV to 30 MeV, common laboratory sources emitting γ-rays up to 4.43 MeV (here: $^{22}$Na, $^{232}$Th, the $^{40}$K background line and $^{241}$Am/$^9$Be) were used. At higher energies γ-rays can be generated by bremsstrahlung, like it was done for the LAT [4], by the use of Compton backscattering, which can be realized inside a storage-ring free electron laser (eg.: HIgS FELL facility [5]) or via (p,γ)-reactions as chosen in our case. For this purpose the small electrostatic Van-de-Graaff accelerator at SLAC, that produces a proton beam up to 350 keV, was reactivated, which was already used to verify the LAT photon effective area at the low end of the GLAST energy range (20 MeV) [6]. Its proton beam strikes a LiF target that terminates the end of the vacuum pipe (see Fig. 1.) and produces 6.1 MeV, 14.6 MeV, and 17.5 MeV gammas via the reactions:

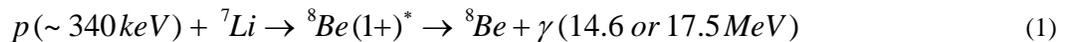

$$p(\sim 340\,keV) + {}^7Li \rightarrow {}^8Be(1+)^* \rightarrow {}^8Be + \gamma\,(14.6\ or\ 17.5\,MeV) \qquad (1)$$

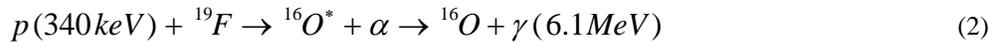

$$p(340\,keV) + {}^{19}F \rightarrow {}^{16}O^* + \alpha \rightarrow {}^{16}O + \gamma\,(6.1\,MeV) \qquad (2)$$

The highly excited 17.6 MeV state of $^8$Be can be created by protons in a resonance capture process at 440 keV on $^7$Li [see reaction (1)]. At lower energies, photons are still produced from the Breit-Wigner tail (Γ =12 keV [9]) of the $^8$Be$^*$ resonance. A narrow γ-ray line at 17.5 MeV is produced by the transition to the $^8$Be ground state, in which the quantum energy is determined by $h\nu = Q + \frac{7}{8}E_p$, with $Q$ = 17.2 MeV as the energy available from the mass change and $E_p$ = 340 keV the proton beam energy. The γ-ray line observed at 14.6 MeV, which corresponds to transitions to the first excited state of $^8$Be, is broadened with respect to the experimental resolution, because of the short lifetime of the state against decay into two alpha-particles. In the resonant case an isotropic flux with a 17.5 MeV/14.6 MeV line intensity ratio of 2:1 is expected [7]. The nonresonant case (eg. for proton energies of ~ 1.15 MeV [8]) is strongly anisotropic and different for the two γ-ray components. When the narrow ($\Gamma$ =3.2 keV [9]) $^{16}$O* resonance at 340 keV is hit, 6.1 MeV γ-rays are generated [see reaction (2)].

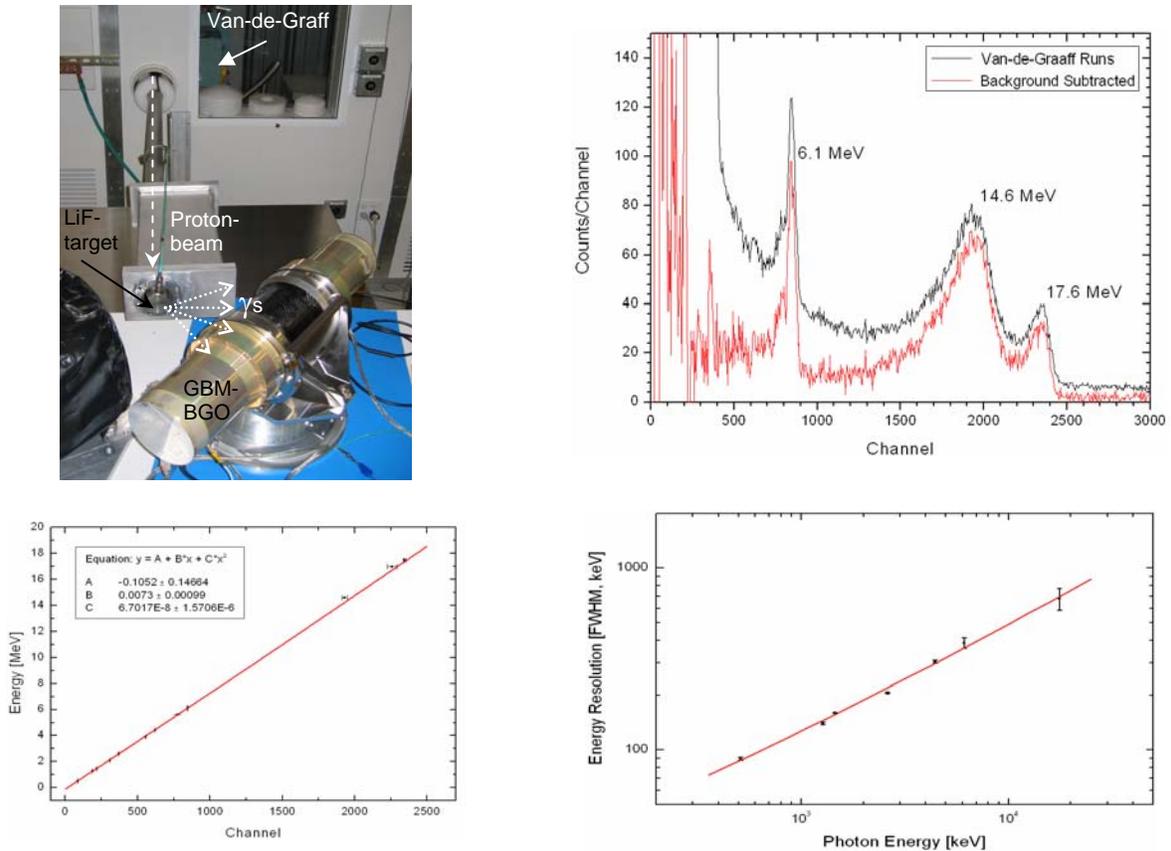

**FIGURE 1.** Upper left: Setup of the GBM-BGO detector at the SLAC Van-de-Graff accelerator. Upper right: Summed spectrum of all Van-de-Graff runs, showing the prominent γ-ray lines of the (p,γ)-reactions (1) and (2). Lower left: Energy to channel calibration. Lower right: Dependence of the energy resolution (FWHM).

The GBM BGO spare detector was placed as close as possible to the LiF-target (Fig. 1, upper left), in order to guarantee a maximized flux of the generated γ-rays, and at an angle of ~45°, with respect to the proton-beam line. The peaks in the high energy-spectrum shown (Fig. 1, upper right) were fitted by accounting for the 511 keV single escape peaks. The photo- to escape-peak ratio was fixed with the help of simulations performed especially for this geometry (single BGO crystal, fan-shaped γ-rays emission region emerging from the location of the LiF-target.). Due to the fact, that the 14.6 MeV line is intrinsically broadened it couldn't be added to the dependence of the energy resolution (Fig. 1, lower right). The energy/cannel relation was best fit by a parabolic function (Fig. 1, lower left). In contrast to the expectations (see above) the measured ratio of the 17.5/14.6 MeV line flux is ~ 1:5, which is not yet understood. The flux ratio was found to be constant when moving the BGO detector to the 0° position.

## REFERENCES

1. A. von Kienlin et al., SPIE 5488, 763 (2004)
2. C. Meegan et al., these proceedings.
3. A. von Kienlin et al., these proceedings.
4. N. Mazziotta et al., these proceedings
5. Litvinenko V. N., et al., SPIE 2521, 55 (1995)
6. Gary Godfrey, LAT-TD-01396-01 (2004)
7. R.L. Walker and B.D. McDaniel, Phys. Rev 74, 315 (1948)
8. M.B. Stearns and B.D. McDaniel, Phys. Rev 82, 450 (1951)
9. T.W. Bonner and J.E. Evans, Phys. Rev. 73, 666 (1948)